\documentclass{aa}  

\usepackage{graphicx, natbib, color, colortbl}
\usepackage{amsmath}
\usepackage{txfonts}
\definecolor{cyan}{rgb}{0.88, 1.0, 1.0}

\begin{document}

   \title{A critical reassessment of the fundamental properties of GJ 504: Chemical composition and age\thanks{Based on observations made with ESO Telescopes at the La Silla Paranal Observatory  under programme ID 072.A-9006(A) and 083.A-9003(A)} }


   \author{V. D'Orazi
          \inst{1,2,3}
          \and
          S. Desidera\inst{1}     
          \and
          R. G. Gratton\inst{1}
          \and
          A. F.  Lanza\inst{4}
          \and
          S. Messina\inst{4}
         \and
           S. M. Andrievsky\inst{5,6}
          \and
          S. Korotin\inst{5,7}
           \and 
         S. Benatti\inst{1}
          \and
          M. Bonnefoy\inst{8, 9}
          \and
          E. Covino\inst{10}
          \and
          M. Janson\inst{11}
          }

   \institute{INAF Osservatorio Astronomico di Padova, vicolo dell'Osservatorio 5, I-35122, Padova, Italy\\
              \email{valentina.dorazi@oapd.inaf.it}
              \and{Department of Physics and Astronomy, Macquarie University, Sydney, NSW 2109, Australia}
              \and{Monash Centre for Astrophysics, School of Physics and Astronomy, Monash University, Melbourne, VIC 3800, Australia}
         \and
                 {INAF Osservatorio Astrofisico di Catania, via S. Sofia 78, I-95123, Catania, Italy}
          \and
         {Department of Astronomy and Astronomical Observatory, Odessa National University, Isaac Newton Institute of Chile,
        Odessa Branch, Shevchenko Park, 65014 Odessa, Ukraine}
         \and
         {GEPI, Observatoire de Paris, PSL Research University, CNRS, Universit\'{e} Paris Diderot, Sorbonne Paris Cit\'{e}, Place Jules Janssen, 92195 Meudon, France}
         \and
         {Crimean Astrophysical Observatory, Nauchny 298409, Republic of Crimea}
         \and
        {Univ. Grenoble Alpes, IPAG, F-38000 Grenoble, France}
         \and
        {CNRS, IPAG, F-38000 Grenoble, France}
         \and
        {INAF Osservatorio Astronomico di Capodimonte, Salita Moiariello 16, I-80131, Napoli, Italy}
         \and
        {Stockholm University, AlbaNova University Center, Stockholm, Sweden}
             }

   \date{Received ; accepted }

 
  \abstract
   {The recent development of brand new observational techniques and theoretical models have greatly advanced the
exoplanet research field. Despite significant achievements, which have allowed the detection of thousands extrasolar
systems, a comprehensive understanding of planetary formation and evolution mechanisms is still desired. One
relevant limitation is given by the accuracy in the measurements of planet-host star ages. The star GJ 504 has
been found to host a substellar companion whose nature is strongly debated. There has been a recent difference of opinion in the literature owing to the uncertainty on the age of the system: a young age of $\sim$ 160 Myr would imply a giant planet as a companion, but a recent revision pointing to a solar age ($\sim$ 4 Gyr) instead suggests  a brown dwarf.}
   {With the aim of shedding light on this debated topic, we have carried out a high-resolution spectroscopic study of GJ 504 to derive 
   stellar parameters, metallicity, and abundances of both light and heavy elements, providing a full chemical characterisation. The main objective is 
   to infer clues on the evolutionary stage (hence the age) of this system.}
   {We performed a strictly differential (line-by-line) analysis of GJ 504 with respect to two reference stars, that is the planet-host dwarf $\iota$ Hor and the subgiant 
   HIP 84827. The former is crucial in this context because its stellar parameters (hence the evolutionary stage) is well constrained from asteroseismic observations.
   Regardless of the zero point offsets, our differential approach allows us to put tight constraints on the age of GJ 504 with respect to $\iota$ Hor, thereby minimising the internal uncertainties.}
   {We found that the surface gravity of GJ 504 is 0.2 $\pm$ 0.07 dex lower than that of the main-sequence star $\iota$ Hor, 
   suggesting a past turn-off evolution for our target. The isochrone comparison provides us with an age range
between 1.8 and 3.5 Gyr, with a most probable age of $\approx$ 2.5 Gyr. Thus, our findings support an old age for the system; further evidence comes from the barium abundance, which is compatible with a solar pattern and not enhanced as observed in young stars.}
   {We envisaged a possible engulfment scenario to reconcile all the age indicators (spectroscopy, isochrones, rotation, and activity); this engulfment could have occurred very recently and could be responsible for the enhanced levels of rotation and chromospheric activity, as previously suggested.
   We tested this hypothesis, exploiting a tidal evolution code and finding that the engulfment of a hot Jupiter, with mass not larger than $\approx$ 3 M$_{j}$ and initially located at $\approx$ 0.03 AU, seems to be a very likely scenario.}


   \maketitle
%

\section{Introduction}
Hunting for planetary systems outside our solar system, with the ultimate goal of revealing habitable worlds, is undoubtedly one of the most intriguing  
and fascinating topics in modern astrophysics. 
Outstanding efforts have been put forwards to detect (and characterise) extrasolar systems, such that 5437 planet candidates have been identified 
so far\footnote{Source \url{http://exoplanets.eu}, as for July 2016}. The presence of those systems has been mostly inferred via indirect methods (e.g. radial velocity variations or transits). However, direct imaging techniques are starting to provide us with  powerful and
complementary tools to detect planetary companions, thanks to the advent of new-generation instruments purposely designed for this goal, such as GPI, ScEXAO,  and SPHERE (see  \citealt{macintosh14}; \citealt{martinache09}; \citealt{beuzit08}, respectively).
Unfortunately, beyond the technical limitations, which are nowadays primarily due to adaptive optics performance, several other fundamental issues affect our comprehension of the substellar regime, 
preventing us from gathering extensive knowledge.
Along with several controversies related to, for instance, planetary atmosphere models, evolutionary stages of substellar objects, different formation, and early evolution scenarios
(core accretion versus disc instability, and/or hot-start versus cold-start models; \citealt{pollack96}; \citealt{kratter10}; \citealt{marley07}; \citealt{fortney08}), the major concern in characterising  exoplanetary systems is represented by the uncertainty on the stellar age. The age estimate of the planet-host star dramatically impacts the inferred mass for substellar companions and hence the calibration of the age-luminosity relationship for substellar objects. Thus, it is crucial to our understanding
of how planets have formed.  

Expanding upon our previous work on GJ 758 (see \citealt{vigan16}), in this paper we present a similar investigation focussing on the curious case of GJ 504. The age of this star is hotly debated. 
The short rotational period of P$_{rot}$=3.3 days (\citealt{donahue96}) and the pronounced level of chromospheric and coronal activities seem to suggest a young age for the system, that is  
160$^{+70}_{-60}$ Myr from gyrochronology and, from the chromospheric level, slightly older at 330$\pm$180 Myr, as traced by the Ca~{\sc ii} H and K emission. 
Assuming the young age of 160 Myr and spectroscopic parameters as published by \cite{valenti05}, \cite{kuzuhara13} concluded that GJ 504 is a young, main-sequence star with $T_{\rm eff}$$\approx $6200 K and log$g$ = 4.6 dex (with $g$ in cm s$^{-2}$). These authors were able to claim that the substellar companion orbiting at 43.5 AU, discovered through direct imaging observations within the framework of the SEEDS survey, has an inferred mass of 4$^{+4.5}_{-1.0}$ M$_{Jup}$. This is the lowest mass companion ever directly imaged to date. 
However, \cite{fuhrmann15} questioned this finding, arguing that the star is actually much older and is comparable with a turn-off stage of evolution and a solar-like age. In addition, these authors 
envisaged a possible merging scenario, with a now engulfed substellar object, to account for the enhanced level of activity and rotation. This study suggested an upward revision for the companion mass, implying that we are dealing with a brown dwarf (M$\approx$30-40 M$_{Jup}$) rather than a giant planet.
An additional motivation for the interest in the GJ 504 system is derived by \cite{skemer16}, who found an indication that the
substellar companion is more enriched ([M/H]$\approx$+0.5) in metal than its parent star ([M/H] $\approx$+0.1-0.3).

We carried out a strictly differential analysis of GJ 504 with respect to three other reference stars ($\iota$ Horologii, HIP 84827, and the Sun) to shed light on this complicated picture, by obtaining in this way very accurate spectroscopic parameters and abundances. We derived temperature, gravity, metallicity, and several key species abundances, from the light  Li and C up to 
elements synthesised via neutron-capture reactions (Ba). 
The manuscript is organised as follows: In Section \ref{sec:abu} we present  the spectroscopic observations and abundance analysis procedure and results. In Section~\ref{sec:age} we extensively discuss all the other age indicators, whereas we sketch a plausible 
engulfment scenario to account for all the observed properties in Section~\ref{sec:merging}. 
We stress that in the present work we are focussing on the fundamental properties of the parent star. The complete characterisation of the substellar companion GJ 504b 
will be published in a companion paper (Bonnefoy et al., in preparation).


\section{Spectroscopic parameters and elemental abundances}\label{sec:abu}

\subsection{Observations and analysis}

We exploited high-resolution (nominal resolution R=48,000), high signal-to-noise FEROS (\citealt{kaufer99}) spectra for all the stars under scrutiny. The observations were carried out during nights in
December 2003 and April 2009 and the original data were retrieved through the ESO archive.
Data reduction was performed using a modified version of the FEROS-DRS pipeline (running under the ESO-MIDAS context FEROS), yielding the wavelength-calibrated spectrum that is merged and normalised through standard steps. The spectra provide an almost complete coverage from 3500 \AA ~to 9200 \AA ~with S/N between 300 and 500 at 6700 \AA~ for the 
three stars; \citet{desidera15} provides further details.
Along with GJ 504, we analysed $\iota$ Hor and HIP 84827 as reference stars. The former is a planet-host, main-sequence star that is relatively young ($\approx$600 Myr) and slightly metal rich
([Fe/H]$\approx$0.15-0.20 dex). It is suspected to be an evaporated member from the Hyades supercluster (\citealt{vauclair08}; \citealt{biazzo12}). 
The activity level and projected rotational velocity of $\iota$ Hor are 
similar to those of GJ 504, ensuring minimal impact of line blending and 
possible differences in the atmospheric structure of active stars (\citealt{santos08}; \citealt{dr09}) 
when performing a differential abundance analysis.  
Conversely, HIP 84827 is an evolved, chromospherically quiet star with metallicity around +0.2 dex and gravity in agreement with a post turn-off evolution (e.g. \citealt{ramirez14}). 
Given the effective temperature, $T_{\rm eff}$ around 6000 K, and the metal-rich nature of $\iota$ Hor and HIP 84827, they are much more similar in terms of atmospheric structural properties to GJ 504 than to the Sun; for this reason, we selected these stars as reference frames for our strictly differential analysis. For sake of completeness, we also analysed a solar spectrum acquired with the same instrument to furnish absolute abundances. The solar values obtained within the present study for species from Li to Ba are reported in Table~\ref{tab:sun}; we adopted $T_{\rm eff, \odot}$=5777 K, log$g_{\odot}$=4.44 dex, $\xi_{\odot}$=0.88 km s$^{-1}$, and A(Fe)$_{\odot}$=7.50 dex.

\begin{table}[h!]
\centering
\caption{Solar abundances from the present study and from \cite{grevesse96}, \cite{asplund09}, and meteoritic abundances from 
\citealt{lodders09} (G+96, A+09, and L+09, respectively).}\label{tab:sun}
\begin{tabular}{lcccr}
\hline\hline
Species & This work & G+96 & A+09 & L+09 \\
\hline
 & & & & \\
 Li & 1.05 & 1.16  & 1.05 & 3.26 \\
C & 8.41        & 8.55 & 8.43 & 7.39    \\
O & 8.63        & 8.97 & 8.69 & 8.40    \\
Na & 6.23        & 6.33 & 6.24 & 6.27 \\
Mg & 7.58        & 7.58 & 7.60 & 7.53   \\
Al  & 6.36       & 6.47 & 6.45 & 6.43   \\
Si & 7.50       & 7.55 & 7.51 & 7.51    \\
S & 7.11        & 7.33 & 7.12 & 7.15    \\
K & 5.07   & 5.12 & 5.03 & 5.08 \\
Ca & 6.30       & 6.36 & 6.34 & 6.29    \\
Ti &    4.92    & 5.02 & 4.95 & 4.91   \\
Cr &    5.59    & 5.67 & 5.64 & 5.64 \\
Fe &    7.50    & 7.50  & 7.50 & 7.45\\
Ni &  6.22 & 6.25 & 6.22 & 6.20 \\
Zn &    4.61    & 4.60  & 4.56 & 4.63\\
Ba & 2.17       & 2.13  & 2.18 & 2.18\\
     &     & & & \\
\hline\hline
\end{tabular}
\end{table}

\begin{figure}
\includegraphics[width=0.4\textwidth]{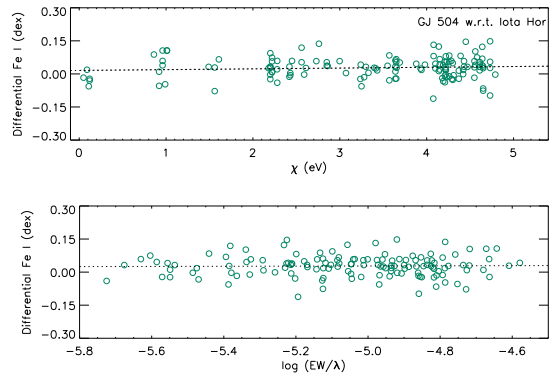}
\caption{Differential iron abundances vs. excitation potential (top panel) and reduced equivalent width (bottom panel) of Fe I lines. The
dashed lines are linear fits to the Fe I lines}\label{f:trends}
\end{figure}

\begin{figure}
\includegraphics[width=0.4\textwidth]{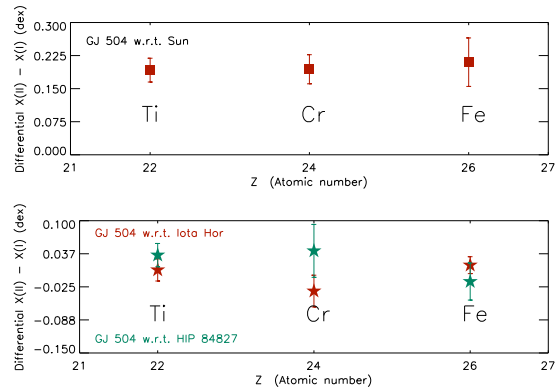}
\caption{Singly ionised minus neutral differential abundances of Fe, Cr, and Ti for differential analysis with respect to the solar spectrum (upper panel), $\iota$ Hor and HIP 84827 (lower panel).}\label{f:ionisation}
\end{figure}

We conducted abundance analyses through equivalent width (EW) measurements for all species but Li, K, and Ba, for which we instead performed spectral synthesis computation.
We measured the EWs with the ARES code (\citealt{sousa07}), but we rigorously double checked every single line with the IRAF\footnote{IRAF is the Image Reduction and Analysis Facility, a general purpose software system for the reduction and analysis of astronomical data. IRAF is written and supported by National Optical Astronomy} task $splot$ with special care to the bluest wavelength domain ($\lambda\lambda$ $\lesssim$ 5000 \AA) , where the severe crowding hampers a straightforward definition of the continuum. Our line-list contains a grand total of 136 and 18 lines for Fe~{\sc i} and Fe~{\sc ii}, respectively, and it is available upon request. For other species we instead selected spectral lines (adopting corresponding atomic parameters) from \cite{melendez14}.
We derived LTE abundances using {\sc moog} by C. Sneden (\citeyear{sneden73}, 2014 version) and employing the Kurucz grid of model atmospheres (\citealt{castelli04}) with solar-scaled chemical composition and no overshooting. 
{\it We obtained both stellar parameters and elemental abundances
by means of a differential line-by-line analysis for GJ 504 with respect to $\iota$ Hor and HIP 84827.}
The first step consists in the determination of the  atmospheric parameters and metallicity by means of the spectroscopic analysis, following the standard procedure.
Effective temperatures ($T_{\rm eff}$) were derived by zeroing the slope between 
differential abundances from Fe~{\sc i} lines and the excitation potential of the spectral features. Similarly, microturbulence values ($\xi$) were obtained imposing no spurious trends between differential abundances from Fe~{\sc i} and the reduced EWs (EW/$\lambda$). Surface gravities (log$g$) instead come from the ionisation balance, that is  $<$$\Delta$(FeII)$-\Delta$(FeI)$>$=0.
The solution is reached when all the three conditions are simultaneously satisfied better than 1$\sigma$ from the error on slopes for temperature and microturbulence and better than roughly one-third the error bar in FeI and FeII features (i.e. the standard deviation from the mean), as previously carried out by \cite{melendez14}. 
We show in Fig.~\ref{f:trends} differential abundances of GJ 504 with respect to  $\iota$ Hor as a function of the excitation potential (upper panel) and the reduced EW (lower panel). 
The gravity optimised by using Fe lines also satisfies the ionisation 
 balance for Cr and Ti, within the observational uncertainties, corroborating our gravity estimate. This is evident from  Fig.~\ref{f:ionisation}, where abundances from neutral and singly ionised lines are shown for Cr, Ti, and Fe. 
 We performed this procedure twice and we assumed as final spectroscopic parameters for GJ 504 the average estimate, for $T_{\rm eff}$, log$g$, $\xi$, and [Fe/H],  with respect to $\iota$ Hor and HIP 84827 (see next section).
Once atmospheric parameters and metallicity are inferred, we proceed to derive the other elemental abundances.
Oxygen abundances,  which come from the EW measurements of the strong permitted triplet at 7771-7775 \AA, were corrected for 3D effects and departures from the LTE, following prescriptions by \cite{amarsi2015} \footnote{\url {http://inspect-stars.com}}.

As previously mentioned, abundances for Li, K, and Ba come from spectral synthesis, employing the driver {\it synth} in {\sc moog}.
As for Li we used the same list as in our previous works (e.g. \citealt{desidera11}), which covers from 6700 to 6720 \AA, including the strong Li~{\sc i} doublet at 6707.78 \AA. 
We calculated  the best match between observed and synthetic spectra, changing the Li abundances and exploiting the 
Ca~{\sc i} line at 6717.69 \AA~ to evaluate the spectral broadening.

In our spectra only two K~{\sc i } lines are available for
the quantitative analysis, that is the resonance doublet at 7665$-$7699 \AA.
The first line is often blended with terrestrial bands. At the solar 
metallicity both lines have very strong wings and this prevents the 
accurate measurements of their equivalent widths. In this case the 
better way to analyse these lines is to use the synthetic spectrum method.

The non local thermodynamical equilibrium (NLTE) effects are rather strong in these lines \citep{bruls92,ivanova00}.
To take these effects into account, we used the atomic model that includes all
transitions between the first 20 levels and the ground level of K~{\sc ii}.
This model was described in details in \citet{andrievski10}.

Barium abundances were derived by synthesing the two Ba~{\sc ii} lines at 5853 \AA~ and 6141 \AA, taking hyperfine structure and isotopic splitting into account  (from \citealt{mcwilliam95}).
The Ba~{\sc ii} feature at 6141 \AA\  is known to be blended with a Fe~{\sc i} line, which is particularly strong in our case given the metal-rich nature of our objects. However, 
the spectral synthesis technique allows us to properly take into account this blend and the strictly differential analysis minimises this source of error.  
Also, we indeed 
derive a solar abundance of A(Ba)=2.21 dex from this feature, in nice agreement with accepted values for solar Ba abundances (see Table~\ref{tab:sun}).
We adopted  an isotopic solar mixture of 81\% ($^{134}$Ba +$^{136}$Ba +$^{138}$Ba) 
and 19 \% ($^{135}$Ba +$^{137}$Ba), as in all our previous studies (e.g. \citealt{desidera11}; \citealt{dorazi12}; \citealt{desilva13}).
In the parameter range of our sample stars, departures from LTE for the Ba~{\sc ii} lines under scrutiny here are very small, thus we did not include them 
(see e.g. \citealt{korotin15}).

\subsubsection*{Error budget} 
External (systematic) and internal (random) errors affect abundance analysis determinations. 
However, given the differential nature of our analysis, for the present purpose we can focus only on internal uncertainties, neglecting zero-point effects of our atmospheric parameters and 
elemental analysis. Two sources of errors can be disentangled: 
\begin{itemize}

\item[] {\it (i) Errors related to the EW measurements} (and/or spectral synthesis) that reflect uncertainties due to the continuum placement and the measurement procedure itself; they can be represented by the standard deviation from the mean (r.m.s), when more than one line for a species is available. 
For elements for which we are forced to rely only on one single feature (K, Zn), we repeated the measurement several times by changing the continuum displacement and other criteria and inspected the corresponding variation in the resulting abundances. 
Typical errors (labelled as observational errors in Table~\ref{tab:errors}) range from 0.018 to 0.105 dex in [X/H] ratios.

\item[] {\it (ii) Errors due to atmospheric parameters.}
In order to assess the contribution related to the stellar parameters, we first need to ascertain the sensitivities of our species to changes in atmospheric quantities. 
To do this, we proceeded in the standard way, by varying one parameter at the time and inspecting the corresponding change in the resulting abundance.
The following step is to evaluate the actual error in atmospheric parameters that originates from errors on the slope for $T_{\rm eff}$ and $\xi$; when estimating 
the uncertainty in gravity we took into account both the standard error from EW measurements of the FeII and the error in $T_{\rm eff}$; that is an error of 20 K would produce a variation in log $g$ of 0.06 dex.
We found values of  20K, 0.07 dex, 0.1 km s$^{-1}$, and 0.04 dex for temperature, gravity, microturbulence, and input metallicity, respectively.
The different contributions are subsequently added in quadrature to gather the total uncertainties due to the stellar parameters (labelled as Param in Table~\ref{tab:errors}).

Given the very (internally) accurate parameter determination, the observational scatter is by far the dominant source of error in our abundances.

\end{itemize}
 
\subsection{Results}\label{sec:abu}

Our results for the two reference stars, $\iota$ Hor and HIP 84827, provide the following atmospheric parameters and metallicity.
We obtained $T_{\rm eff}$=6230$\pm$60 K, log$g$=4.45$\pm$0.15 dex, $\xi$=1.06$\pm$0.10 km s$^{-1}$ and [Fe/H]=0.19$\pm$0.06 dex and $T_{\rm eff}$=6155$\pm$60 K, log$g$=4.15$\pm$0.15 dex, $\xi$=1.00$\pm$0.10 km~s$^{-1}$ and 
[Fe/H]=0.19$\pm$0.08 dex for $\iota$ Hor and HIP 84827, respectively.
We would like to emphasise that both estimates are in very good agreement with previous determinations of parameters and abundances for the two reference stars, suggesting that systematic uncertainties do not largely affect our analysis
(see for example \citealt{biazzo12} and \citealt{ramirez14}). More importantly, our parameter determination for the planet-host star $\iota$ Hor agrees within 1 $\sigma$ with the asteroseismic results (see \citealt{vauclair08}).

The differential (line-by-line) spectroscopic analysis provides the following differential parameters for GJ 504 when $\iota$ Hor is taken as reference: d$T_{\rm eff}$=$- 40\pm$20 K, dlog$g$=$-0.20\pm$0.07 dex, d$\xi$=+0.15$\pm$0.10 km s$^{-1}$ and 
d[Fe/H]=+0.03$\pm$0.03 dex. On the other hand, we found d$T_{\rm eff}$=+65$\pm$20 K, dlog$g$=+0.17$\pm$0.07 dex, d$\xi=+0.24\pm$0.10 km s$^{-1}$, and 
d[Fe/H]=+0.03$\pm$0.04 dex when the analysis is differential with respect to HIP 84827. We then calculated the absolute abundances for GJ 504 relative to these reference stars and we finally took the average values as our adopted stellar parameters and metallicity, that is
$T_{\rm eff}$=6205$\pm$20 K, log$g$=4.29$\pm$0.07 dex, $\xi$=1.23$\pm$0.10 km s$^{-1}$ and 
[Fe/H]=0.22$\pm$0.04 dex. 
In Table~\ref{tab:comparison} we list for GJ 504, along with our stellar parameters and [Fe/H], literature values from several recent studies, based on high-resolution spectroscopy.
The temperature is in good agreement with previous estimates as well as metallicity and microturbulence values; 
we confirm that the star is characterised by a super-solar metallicity.
The crucial result here is represented by the surface gravity.
Our analysis (as the vast majority of previous spectroscopic studies with the exception of \citealt{valenti05} and \citealt{maldonado12}) indicates 
that the star is actually slightly evolved, and compatible with the turn-off evolutionary stage. 
We stress that the agreement between the temperature value from the present work and \cite{valenti05}, but a strong mismatch in gravity, is not surprising because in their paper gravities have been evaluated from the fitting of the Mg b lines; 
hence they are uncorrelated with $T_{\rm eff}$.
Thus our result is in contrast to what has been suggested by \cite{kuzuhara13}, who adopted gravity from \cite{valenti05} (i.e. young age), 
corroborating the conclusions of Fuhrmann and Chini. The implications of our results in terms of age are discussed in the next Section and are crucial when interpreting 
the substellar nature of the companion (see also the introduction of this paper).

At first glance, other element abundances track the iron content reasonably well. A closer inspection reveals, however, that GJ 504 exhibits a small overabundance in K, which is  
[K/Fe]=0.15 $\pm $0.08 dex. When observational errors are taken into account this value is still compatible with a solar-scaled pattern.
Moreover, carbon and oxygen are underabundant. We recall that since NLTE corrections have been applied, our results do not depend on the wrong assumption of local thermodynamical equilibrium. Had not we applied this corrections, the [K/Fe] ratios would have been $\approx$ 0.50 dex. The same holds for oxygen abundances,
for which NLTE corrections are about  $-$0.3 dex. It is noteworthy in this context, that \cite{bensby2014} 
obtained a [O/H]=0.11$\pm$0.11 dex (thus subsolar [O/Fe] ratio), which is in agreement with our value once errors are considered. 
Besides, the impact of NLTE departures is actually much smaller when we consider differential abundances of GJ 504 with respect to $\iota$ Hor because the stars are extremely similar in terms of atmospheric parameters with the clear exception of gravity
(where $\Delta$ log$g$=0.20$\pm$0.07 dex). 

The Li abundance is discussed in the next section, which is dedicated to a reassessment of all the other age indicators. 
The Li content is indeed known to be a critical key diagnostics for stellar ages of relatively cool stars, 
as shown by numerous studies (e.g. \citealt{sestito05}).
Here we mention that we obtained A(Li)=2.88$\pm$0.07, from the spectral synthesis of the Li doublet. 
This value agrees very well with previous analyses by \cite{takeda05} and \cite{ramirez2013}, who obtained A(Li)=2.91$\pm$0.10 and A(Li)=2.85$\pm$0.04, respectively.

\begin{table*}
\centering
\caption{Comparison between GJ 504  stellar parameters and metallicity from this work and several, recent literature values.}\label{tab:comparison}
\begin{tabular}{lcccr}
\hline \hline
$T_{\rm eff}$ & log$g$ & $\xi$ & [Fe/H] & Reference \\
     (K)           & (dex) & (kms$^{-1}$)  & (dex) &  \\
 \hline                   
 &             &         &               &        \\
 \rowcolor{cyan}
 6205$\pm$20   & 4.29$\pm$0.07 & 1.23$\pm$0.10 & 0.22$\pm$0.04 & This work \\
 5978$\pm$60  &  4.23$\pm$0.10 & 1.13$\pm$0.20 & 0.13$\pm$0.06 & Fuhrmann \& Chini (2015) \\
 6234$\pm$25 & 4.60$\pm$0.02 &         --------                   & 0.28$\pm$0.03 & \cite{valenti05}\\ 
 6012$\pm$100  & 4.30$\pm$0.20 & 1.10$\pm$0.20 & 0.11$\pm$0.10 & \cite{mishenina2013}\\ 
 6185$\pm$51  & 4.30$\pm$0.07  & 1.28$\pm$0.08 & 0.25$\pm$0.06 & \cite{battistini2015}\\
 5995$\pm$41     &  4.24$\pm$0.02          &  1.34$\pm$0.12                 & 0.11$\pm$0.04                   & \cite{ramirez2013}\\      
 6133$\pm$50 & 4.63$\pm$0.13 & 1.30$\pm$0.26 & 0.24$\pm$0.04 & \cite{maldonado12}\\
 & & & & \\
 \hline\hline
\end{tabular}
\end{table*}

\begin{table}
\centering
{\renewcommand{\arraystretch}{1.2}
\renewcommand{\tabcolsep}{0.1cm}
\caption{Differential abundances of GJ 504 with respect to $\iota$ Hor and HIP 84827 in Cols. (2) and (3), respectively. Oxygen has been corrected for 3D+NLTE effects, following prescriptions by \cite{amarsi2015}. Similarly we provide NLTE K abundances. Listed values for Ti, Cr, and Fe are the average values between singly ionised and neutral lines.}\label{t: abu}
\begin{tabular}{lcccc}
\hline\hline
Elem. & d[X/H]                         &  Obs.             &  d[X/H]                                    & Obs.  \\ 
              & \tiny {(w.r.t $\iota$ Hor)}           &                          &      \tiny {(w.r.t HIP 84827)}     &                   \\
\hline              
C       &        $-$0.057  & 0.083  & ~~0.000 &0.082                 \\
O       &        $-$0.031  & 0.063  & ~~0.050 &0.030         \\
Na      &         ~~0.019  & 0.110  &$-$0.066 &0.101       \\
Mg      &         ~~0.030  & 0.027  &$-$0.083 &0.087       \\
Al      &         ~~0.023   & 0.082 & ~~0.120 &0.031       \\
Si      &         ~~0.067   & 0.058 & ~~0.008 &0.065       \\
S       &        $-$0.099  & 0.077  &$-$0.023 &0.076       \\
K      &       $-$0.010 &  0.070  & ~~0.120 & 0.070        \\
Ca      &         ~~0.046       & 0.056  &~~0.080 &0.047      \\
Ti      &         ~~0.019  & 0.016  & ~~0.015 &0.053       \\
Cr      &        $-$0.028 & 0.021  &$-$0.020 &0.048       \\
Fe      &         ~~0.009 & 0.026  &~~0.011 &0.065        \\
Ni      &         ~~0.060 & 0.065  &$-$0.029 &0.063         \\
Zn      &         ~~0.035 & 0.028  &$-$0.073 &0.045         \\     
        &                                       &                                       &                                               &                       \\
  \hline
  \hline
  \end{tabular}}
  \end{table}
  
  \begin{table*}
\centering
\caption{[X/H] ratios for GJ 504 with respect to the Sun and errors (see text).}\label{tab:errors}
\begin{tabular}{lcccccccr}
\hline\hline
Element & [X/H] & $\Delta T_{\rm eff}$ & $\Delta$log$g$ & $\Delta\xi$ & $\Delta$[M/H] & Param & Observational & Total \\ 
             &           &      (+20 K)&       (+0.07 dex)      &    (+0.10 km s$^{-1}$)    &       (+0.04 dex)         &             &                   &          \\
 \hline
       & & & & & & & & \\            
 C     & $-$0.004     & ~~0.008   & $-$0.027    & $-$0.005   &   $-$0.008  &   0.026  &   0.106   &   0.109            \\
 O     & ~~0.030      & ~~0.024   & $-$0.010    & ~~0.029    &   ~~0.015   &   0.043  &   0.040   &   0.059       \\
 Na    & ~~0.156      & ~~0.154   & ~~0.158     & ~~0.156   &   ~~0.158    &   0.003  &    0.036   &   0.036      \\
 Mg    &  ~~0.201     & ~~0.197   & ~~0.211     & ~~0.199   &  ~~0.201     &   0.011  &    0.060   &   0.061      \\
 Al    & ~~0.187      & ~~0.184   & ~~0.193     & ~~0.185    &   ~~0.187   &   0.007  &   0.018   &   0.019       \\
 Si    & ~~0.215      &~~0.216    & ~~0.212     & ~~0.213    &  ~~0.213    &   0.004  &   0.090   &   0.090       \\
 S     & ~~0.000      &$-$0.003   & $-$0.021    &$-$0.002    &  $-$0.003   &   0.022  &   0.024   &   0.033       \\
K     &  ~~0.370     & ~~0.010    &    $-$0.010       &$-$0.030  &   ~~0.000   &   0.030  &   0.070  &    0.080    \\
 Ca    & ~~0.268      &  ~~0.264  & ~~0.283     &~~0.263    &  ~~0.267     &   0.016  &   0.105   &   0.106        \\
 Ti    & ~~0.192      & ~~0.189   & ~~0.173     &~~0.190     &   ~~0.191   &   0.019  &   0.027   &   0.033       \\      
 Cr    & ~~0.194      & ~~0.192   & ~~0.176     &~~0.190    &  ~~0.193     &   0.019  &   0.033   &   0.038       \\
 Fe    & ~~0.210      & ~~0.222   & ~~0.213     &~~0.199    & ~~0.225      &   0.022  &   0.055   &   0.059       \\
 Ni    & ~~0.233      & ~~0.232   & ~~0.234     &~~0.231     &  ~~0.231    &   0.003  &   0.056   &   0.056       \\      
 Zn    & ~~0.209      & ~~0.210   & ~~0.209     &~~0.195    &  ~~0.203     &   0.015  &   0.050   &   0.052       \\
 & & &  & & & & &                                                                                                \\
 \hline\hline                   
\end{tabular}
\end{table*}

Finally, the barium abundance deserves a special, dedicated discussion. 
In 2009, we discovered that there is a negative correlation between the Ba content and the open cluster age (\citealt{dorazi09}). 
This finding was confirmed by other authors (e.g. \citealt{jacobson13}), providing further evidence that the younger the cluster, the higher the Ba abundance.
Pre-main-sequence clusters, such as IC 2602 and IC 2391 (age between 30 and 50 Myr), exhibit [Ba/Fe] ratios up to $\approx$0.65 dex, that is more than a factor of four the solar value.
Since then, a variety of studies have been conducted to ascertain the nature of this peculiar trend, which does not seem to be accompanied by a similar behaviour from the other elements produced in slow neutron-capture reactions (La in particular, which belongs with Ba to the second peak of the $s$-process path). 
Regardless the nature of such a special chemical pattern (see an extensive discussion in \citealt{dorazi12}), we can exploit this overabundance as a youth indicator.
Crucial to the present work, super-solar Ba ratios in young stars are not confined to the cluster environment, but it has also been shown to be present also in young, isolated stars
(see e.g. HD 61005 by \citealt{desidera11}).
Our analysis indicates that GJ504 is characterised by a solar Ba abundance, which is [Ba/Fe]=$-$0.04$\pm$0.01$\pm$0.03 dex; 
the second error is from stellar parameters, whereas the first is simply the standard deviation from the two different Ba features. 
On a differential scale with respect to $\iota$ Hor, GJ 504 has $\Delta$[Ba/Fe]=$-$0.28$\pm$0.05 dex, implying 
an enhanced Ba content for the reference star, at variance with the allegedly young age of GJ 504. Although based on a different analysis procedure, we confirm that $\iota$ Hor is rich in Ba, as previously determined 
in \cite{dorazi12} and in nice agreement with expectations, according to the young age. 
This is a very important result because the age and the evolutionary stage 
of $\iota$ Hor are very well constrained, given the asteroseismic measurements 
published by \cite{vauclair08}.
Thus, we can conclude that our analysis for Ba abundance determination (along with the gravity) also points to an old age for GJ 504.
%
%

\section{Other age indicators}
\label{sec:age}

\subsection*{Rotational period, chromospheric activity and X-ray emission}

The rotation period P=3.33 d is well established
from long-term Ca II H\&K time series
\citep{donahue96} and photometry \citep{messinathesis}. 
\citeauthor{donahue96} also detected rotation period variations in the range from 
3.23\,d to 3.41\,d that they attributed to the presence of surface differential rotation. 

\begin{figure*} 
\includegraphics[width=0.85\textwidth]{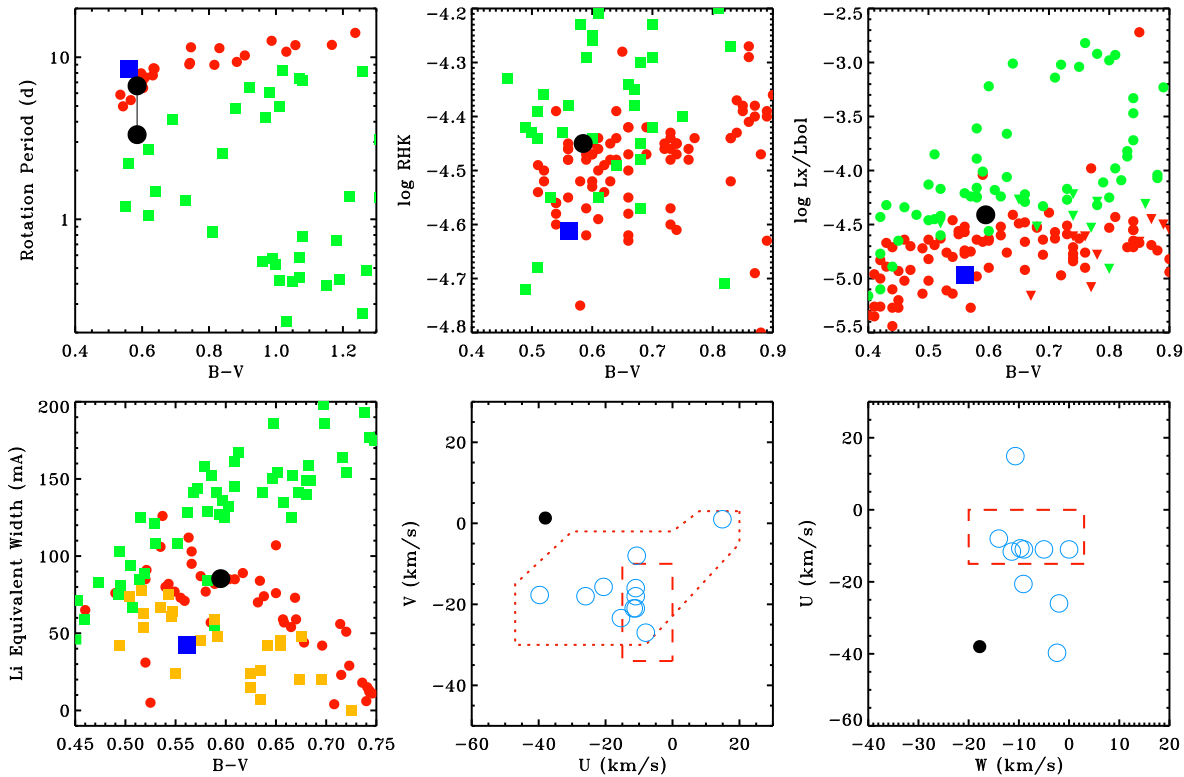}
\caption{Age indicators for GJ 504 compared to those
of $\iota$ Hor and open clusters of well-determined age.
In the various panels, GJ 504  is shown as a filled black circle,
$\iota$ Hor as a filled blue square, members of the Pleiades open cluster  as
green filled squares, members of the Hyades open cluster  as red 
circles, and members of the NGC 752 open cluster as orange filled squares.
Upper left panel: Rotation period vs. B-V. For GJ 504, both
the literature rotation period (3.33d) and the possible alternative
period that is two times longer are plotted and connected by a solid line.
Upper middle panel: $\log R^{'}_{HK}$ vs. B-V. 
Upper right panel: $\log L_{X}/L_{bol}$ vs. B-V.  
In these plots, red filled triangles and green filled triangles
represent the stars with X-ray upper limits members of Hyades and 
Pleiades, respectively.
Lower left panel: Lithium EW vs. B-V for GJ 504 (black square).
Lower central panel: U vs. V space velocities for GJ 504 and young
moving groups (open circles). Overplotted are the boundaries
of the kinematic space populated by young stars.
Lower right panel: W vs. U space velocities for GJ 504  and young
moving groups (open circles).}
\label{fig:prot}
\end{figure*}

This rotation period of 3.33 d is intermediate between those of stars of
similar colours of the Hyades and Pleiades open clusters 
(Fig.~\ref{fig:prot}).
The stellar radius as derived in Sect.~\ref{sec:isoc} allows us to estimate the
inclination at which the star is seen by the Earth.
From the rotation period of 3.33 days and the projected rotational
velocity vsin$i$ = 6$\pm$1 km~s$^{-1}$ (calculated as a by-product of the spectral synthesis method; see also e.g. \citealt{gonzalez10}), we infer
a system inclination of 18 deg.
The nearly pole-on inclination does not prevent the detectability
of photometric rotational modulations, as discussed in 
\citet{messina16}.
It is possible that the true rotation period is two times the proposed one, 
with the shorter periodicity resulting from two roughly symmetric active
latitudes. As the original datasets are not available, we cannot 
conclusively support or discard this possibility, but we 
emphasise that  a period of 6.6 days would not solve the discrepancy
with ages coming from isochrones and spectroscopy. 
Thus our conclusions are unaffected. 

The average index $\log R^{'}_{HK}=-4.446$ \citep{baliunas96} is on the higher envelope   
of Hyades members of same mass (Fig.~\ref{fig:prot}). The 
$\log R^{'}_{HK}$ given in \citet{baliunas96} is slightly different
(-4.443) because of the slightly different B-V colour they adopted.
We rederived the value of $\log R^{'}_{HK}$ for our adopted colour B-V=0.585
for consistency with the other indicators.
The \citet{baliunas96} $\log R^{'}_{HK}$ is based
on a very extended time series of several hundreds of measurements spanning 
20 years, then effectively averaging variability on daily and yearly
timescales. Several other literature sources quote $\log R^{'}_{HK}$ for GJ 504,
mostly based on sparse measurements. They are consistent with the
\citet{baliunas96} results.
The coronal activity, as measured by ROSAT, is also consistent with
the fast rotation and high activity level of the star (Fig.~\ref{fig:prot}).
A photometric time series that is 13 yr long and an HK time series that is 20 yr long show       
that  the  photometric  and chromospheric variability
are found to be anti-correlated with  a significance level $>95$\%.  
This behaviour is typical of young active stars (see e.g.  \citealt{lockwood07}, their Fig. 3).
The estimated stellar ages derived from rotation, $\log R^{'}_{HK}$, and 
$\log L_{X}/L_{bol}$ are 163 Myr, 440 Myr, and 431 Myr, respectively,
when using \cite{mamajek08} calibrations.
In summary, the age indicators linked to stellar rotation indicate an age 
intermediate between the Pleiades and the Hyades.
The rotational and activity parameters of $\iota$ Hor are $P_{rot}=8.5$~d; 
$\log R^{'}_{HK}$= $-$4.61  and  $\log L_{X}/L_{bol}$=$-$4.97 (\citealt{metcalfe10}). 
Therefore, $\iota$ Hor results to be slightly older than GJ 504 from
the rotation/activity point of view and with a high accuracy age 
of 625$\pm$5 Myr from asteroseismology \citep{vauclair08}, further 
supporting our choice as reference young star in the 
differential abundance analysis.

\subsection*{Lithium abundance}

The lithium content of the stellar atmosphere is another key age 
diagnostic, although it is not as sensitive as for cooler stars for 
effective temperature as that of GJ 504.
To avoid the challenges due to possible systematic differences in
temperature scale we compared the equivalent width of Li line
at 6707\AA~~of GJ 504 (and $\iota$ Hor) with those of members of
open clusters of different ages (Pleiades, Hyades, and NGC 752).
Lithium data were taken from \citet{soderblom93}, \citet{thorburn93}, 
and \citet{sestito04}
for Pleiades, Hyades, and NGC 752, respectively.
As shown in 
Fig.~\ref{fig:prot},
the Li EW of GJ 504 lies on the locus of
Hyades members, then with a marginal discrepancy with respect to the younger age inferred from rotation, and clearly above
the locus of the 2 Gyr-old open cluster NGC 752. GJ 504 has a temperature that puts it on the cool border of the Li dip. 
Therefore, its Li content could be slightly larger than observed and in better agreement with an age younger than the Hyades.

\subsection*{Kinematic parameters}
\label{sec:kin}

GJ 504 is not associated with any known young moving group. 
The heliocentric space velocity components U,V,W are found to be $-$38.0$\pm$0.2, +1.3$\pm$0.2, and $-$17.8$\pm$0.5 km\,s$^{-1}$,
respectively, when adopting the proper motion and trigonometric parallax from van Leeuwen (2007) 
and the absolute radial velocity (RV=$-$27 km/s) from \citet{nidever02}.
These parameters are slightly outside the kinematic space of young stars
identified by \citet{montes01}, making an age as young as 
200-300 Myr unlikely.
The BANYAN II on-line tool \citep{gagne14} also yields 
a much higher probability that the star belongs to the old field
rather than the young field (97\% versus 3\%).
The galactic orbit derived by \citet{marsakov95} 
($R_{p}=7.98$ kpc; $R_{a}=10.00$ kpc; $Z_{max}=0.128$ kpc, and $e=0.112$)
further rules out a very young age but it is still
in fair agreement with an age that is a bit younger than the Sun, as argued by
the isochrone fitting in Sect.~\ref{sec:isoc}.

\subsection*{Isochrones}
\label{sec:isoc}

We derived the stellar ages from the isochrone fitting using both
the observed colours and magnitudes and the spectroscopic
parameters from Sect.~\ref{sec:abu}.
We used the PARSEC models by \citet{bressan12} for the appropriate
metallicity of the star (Z=0.0235, corresponding to 
[M/H]=+0.22)\footnote{http://stev.oapd.inaf.it/cgi-bin/cmd, PARSEC version 1.2S}, adopting Z$_{\odot}$=0.0152 as for PARSEC isochrones; see \citealt{bressan12}.

\begin{figure*} 
\includegraphics[width=0.85\textwidth]{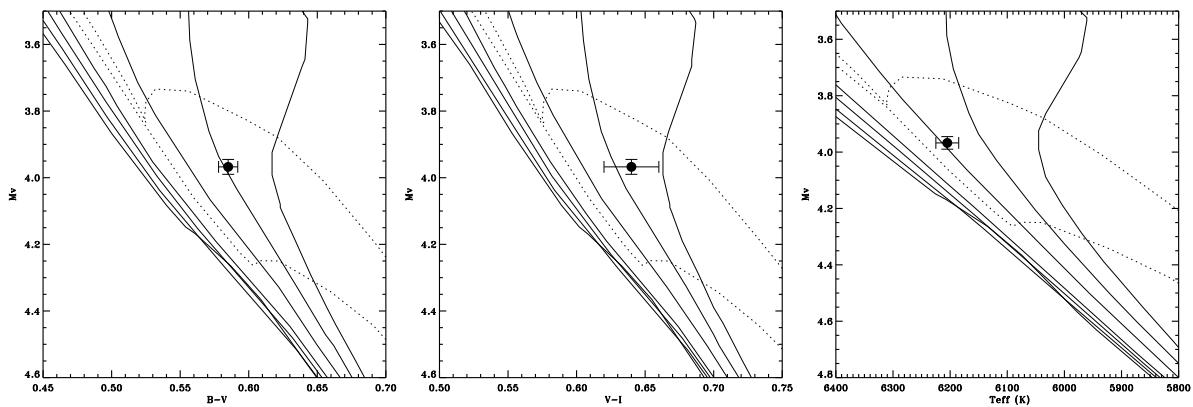}
\caption{$M_{V}$ vs. B-V (left panel), $M_{V}$ vs. V-I (central panel), and $M_{V}$ vs. $T_{\rm eff}$ (right panel)
for GJ 504. Overplotted are the 0.1, 0.4, 0.7, 1.0, 2.0, 3.0, and 4.0 Gyr
isochrones (continuous lines) and the 20 and 25 Myr pre-main-sequence isochrones (dashed lines)
for the appropriate metallicity, from \citet{bressan12}.}
\label{fig:bvvi}
\end{figure*}

Fig.~\ref{fig:bvvi} shows the colour magnitude diagrams using V 
magnitude and B-V and V-I colours. GJ 504 lies about 0.3 mag above 
the ZAMS, close to the 3.0 and 3.2 Gyr isochrones, respectively.
When using the $T_{\rm eff}$ versus $M_{V}$ diagram, GJ 504 appears marginally
younger (age about 2.1 Gyr).
The age as resulting from the $\log g$ versus $T_{\rm eff}$ diagram is 3.0 Gyr.

If the position above main sequence is instead interpreted as a pre-main-sequence evolutionary
phase, the resulting isochrone age is of about 20-25 Myr. Such an extremely young age
is definitely not consistent with the observed lithium content and the kinematic parameters
(space velocities very far from the typical locus of stars younger than 100 Myr).

\begin{figure} 
\includegraphics[width=0.4\textwidth]{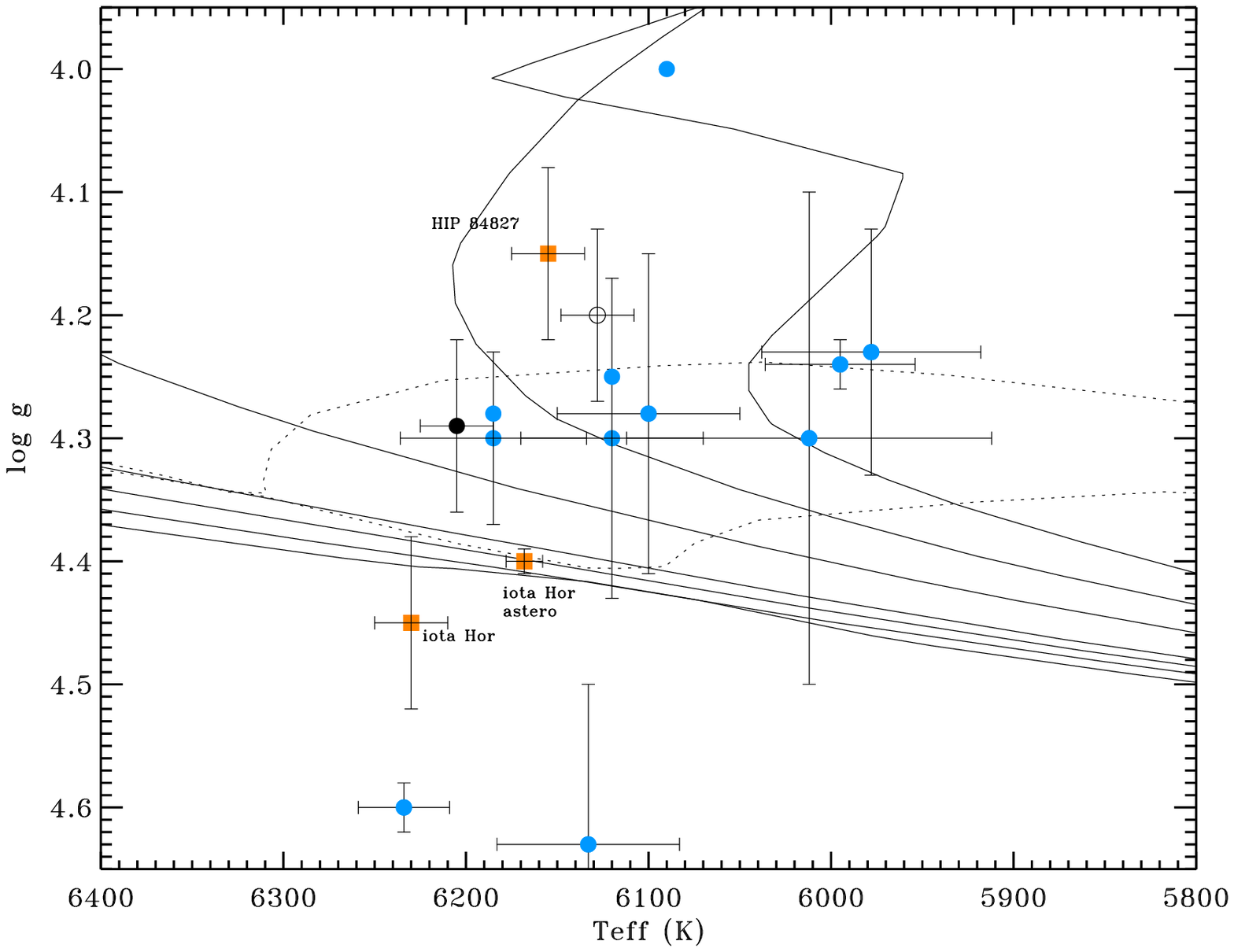}
\caption{$\log g$ vs. $T_{\rm eff}$ 
for GJ 504 as resulting from our spectroscopic analysis (filled black circle) and when applying the
results of the spectroscopic differential analysis with respect to the asteroseismology value
of $\iota$ Hor (empty circle).
Overplotted are the 0.1, 0.4, 0.7, 1.0, 2.0, 3.0, and 4.0 Gyr
isochrones (continuous lines) and the 20 and 25 Myr pre-main-sequence isochrones (dashed lines)
for the appropriate metallicity from \citet{bressan12}.
Additional measurements of $\log g$ vs. $T_{\rm eff}$
for GJ 504 from the literature are plotted as light blue filled circles.
The parameters of $\iota$ Hor and HIP 84827 (our spectroscopic analysis and the
parameters of iota Hor as derived from the asteroseismology analysis
\citep{vauclair08} are plotted as empty red circles.}
\label{fig:tefflogg}
\end{figure}
We then conclude that the isochrone age of GJ 504 is between 1.8 and 3.5 Gyr
with a most probable value of  2.5 Gyr. The corresponding stellar radii and masses are $1.30~R_{\odot}$ and
$1.22~M_{\odot}$, respectively. 
We stress that the absolute age determination of GJ 504 is beyond the scope of the present paper and this will be addressed in a forthcoming work
that will focus on asteroseismic observations of the system. For our purposes only differential properties of GJ 504 with respect to the reference stars
(especially $\iota$ Hor) are relevant.

Our estimate of the isochronal  age indicates that GJ 504 is  slightly younger than that derived by \citet{fuhrmann15}. 
This is mostly because of the different values of effective
temperature adopted. 
However, the discrepancy of isochrone age
with other age indicators such as rotation, activity and lithium
is maintained and calls for a dedicated explanation.
The isochrone age is instead consistent with the kinematic parameters
and the barium abundance.

\subsection*{Age summary}
\label{sec:agesummary}

We explored whether the isochrone age, as resulting from
the evolved position in the colour-magnitude diagram, could be 
spurious as a result of an unresolved binarity of GJ 504.

There are no indications of additional stellar components contributing to 
the integrated flux either from direct imaging (ruling out equal-luminosity 
companions down to a separation of 30-50 mas; Bonnefoy et al., in prep.), 
or from spectroscopy. Radial velocity (RV) data from \citealt{fischer14}),  
direct inspection of the FEROS spectra, and the check for possible 
dependencies of the results of the  abundance analysis on wavelength 
of the spectral lines seems to exclude this scenario. 
Only very special orbit configurations (wide orbits with extremely small
projected separation at the time of the imaging observations) are not yet
ruled out by available observations, however their occurrence probability is vanishingly 
small because of tight observational constraints.
In any case a blended binary scenario would
not easily explain the fully consistent result between the spectroscopic
gravity and the position on colour-magnitude diagram.
The isochrone age is also supported by the kinematic parameters and
barium abundance and any ad-hoc explanation for a spurious estimate
would not remove the discrepancy for these indicators.
Therefore, we dismiss the possibility that the isochrone age is biased
by an unresolved companion and look for alternative explanations
for the discrepancy with the other indicators (rotation, activity, and lithium), 
assuming that the isochrone age is the most reliable for GJ 504.


\section{Solving the age discrepancy: merging events}
\label{sec:merging}

\subsection{Ruling out stellar companions and their remnants}

The discrepancy between the age indicators linked to rotation
and the placement on HR diagram can be ascribed to the 
interaction with a companion, which could spin up the central star.
This could happen through tidal interactions by a close companion,
a merging, or wind accretion by a moderately wide companion
at the end of the AGB phase.

The presence of a massive companion close enough to tidally spin up
the central star is ruled out by the RV monitoring performed at Lick
Observatory \citep{fischer14}. The RV time series include 58 measurements
over 21.65 years. The observed RV dispersion (25.7 m~s$^{-1}$) is fairly large
but comparable with the high activity level of the star.
This allows us to exclude the presence of close companions more massive
than about $0.5-1~M_{J}$. 

The lack of any long-term RV slope also argues against the presence of
a white dwarf (WD) companion at a separation small enough to have allowed
significant accretion of material and angular momentum onto the central
star \citep[see][ for details and references on this scenario]{zurlo13}.
Such a white dwarf companion is also ruled out by the imaging observations.
Finally, a hot WD would be expected if a recent accretion event had occurred 
(as expected from the fast rotation of GJ 504), 
while the far and near ultra violet magnitudes observed by GALEX  are
fully consistent with those of a chromospherically active
G0V star \citep{findeisen11} and clearly rule out a spatially unresolved hot WD
companion to GJ 504. 

This leaves the occurrence of a merging event as the most viable
possibility to account for the various observational constraints,
as already proposed by \cite{fuhrmann15}. 
Following \cite{fuhrmann15}, it should be noticed that the high lithium
content allows us to constrain the mass of the merged object.
Indeed, the merging with a low-mass star of 0.1-0.2 $M_{\odot}$ would
imply that the material accreted is completely depleted in lithium
and an original mass of 1.0-1.1  $M_{\odot}$ for the primary would 
imply some Li depletion after 2-3 Gyr.
The observed Li EW of GJ 504 is instead above 
the locus of the 1.5-2.0 Gyr old NGC 752 open cluster. This is not consistent
with the original Li abundance of a star with mass significantly below
the present one and the accretion of a significant amount of 
fully Li depleted material.
This suggests that the original mass of the primary was very close to the present one,
and that the mass of the merged companion was substellar and most likely low enough to have preserved lithium. 
From the observed mass distribution of close-in substellar companions (\citealt{reggiani16}, and references therein), 
it is more likely to face  a planet rather than a brown dwarf.
We briefly mention, in this context, that the possible difference in lithium content for stars with and without planets is a very debated topic and 
no final conclusion has been drawn yet.  \cite{delgadomena14} and references therein, for example provide for a dedicated 
discussion;  however, the temperature of GJ 504 is hotter than the range ($T_{\rm eff}$$\approx$ 5600 - 5900 K), where 
evidence for a lower Li abundance in planet-host stars has been observed.
Moreover, it is clear that GJ 504 belongs to a completely different class of objects because the central star has already cannibalised the close-in giant planet and we do not 
know what kind of planetary system was present at the beginning of its evolution. On the other hand, all correlations between Li and planet hosts are focussed on 
stars currently orbited by planetary-mass companions,  hence engulfment events did not occur.

We discuss the scenario of the engulfment of a planetary companion in more detail in the next subsection.

\subsection{A possible engulfment scenario of a planetary companion}

To quantitatively evaluate the timescales involved in our proposed engulfment scenario, 
we employed a simple tidal evolution model. It assumes circular and coplanar orbits 
and includes the rate of angular momentum loss produced by the stellar magnetised wind 
$\dot{L}_{\rm w}$ with a Skumanich-type law; that is $\dot{L}_{\rm w} = - K \Omega^{3}$, 
where $K = 3.52 \times 10^{40}$ kgm$^{2}$~s and $\Omega$ is the stellar angular velocity. 
The tidal torque between the star and the planet follows the equilibrium-tide 
formalism (see e.g. \citealt{zahn2008}) with its strength  parameterised by the 
tidal modified quality factor $Q^{\prime}$ of the star. We take into account the 
evolution of the radius of the  star along the main sequence\footnote{We have used 
the EZWeb interface to compute and read a simple stellar
evolution model (\url{http://www.astro.wisc.edu/~townsend/static.php?ref=ez-web)})}. 
The adopted parameters for stellar mass and metallicity are $M=1.2$ M$_\odot$ and  $Z$=0.02,  
respectively. The validity of the Skumanich law implies that we cannot model phases 
with the star rotating  faster than $\approx 5$ days. 
Following \citet{fuhrmann15}, we adopt a mass of $2.7$ M$_j$ for the planet. 

Since the planet is falling into the star, the numerical solution of our tidal 
equations is highly sensitive to the initial conditions when we integrate  them 
forwards in time. To avoid such an instability, we start from the final condition, 
that is with the planet filling its Roche lobe at $a_{\rm f} = 0.009$ ~AU, and integrate 
backwards in time to find the initial condition.  Our model does not take into account 
the mass transfer that occurs when the planet fills its Roche lobe. This phase has been 
modelled by, for example \citeauthor{valsecchi14} (\citeyear{valsecchi14}, \citeyear{valsecchi15}) and \citet{jackson16}, 
who found that its duration depends in a critical 
way on the interplay between the planet mass loss, which tends to increase the semi-major axis, 
and the stellar tides, which shrink it. We also note the above-mentioned publications  concerning evaporation issues.
A complete treatment of this topic is beyond the scope of the present paper and does not affect our conclusions. 
If the planet reaches its Roche lobe early, when 
the star is  on the main sequence and tides are not particularly effective, the mass 
transfer phase can last for hundreds of Myr or Gyr. In our case, we assume that the 
planet loses most of its mass in a relatively short time (tens or hundreds Myr) because tides 
are rapidly shrinking its orbit owing to the radius expansion of our $1.2$~M$_{\odot}$ star 
during the final phase of its main-sequence evolution. Therefore, only the core of the 
planet, if any, should survive the rapid evolution after the onset of Roche lobe overflow. 

 \begin{center}
\begin{figure} 
\includegraphics[width=0.4\textwidth]{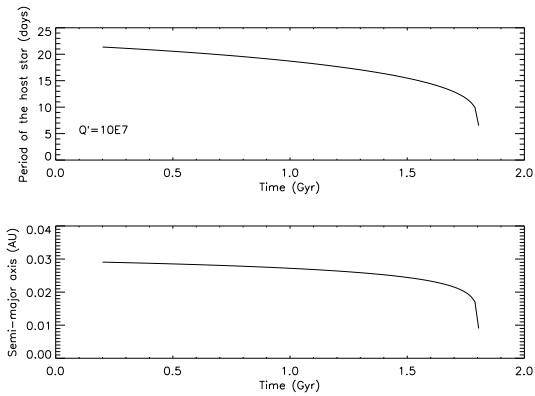}
\caption{Time evolution of the rotational period of the star (upper panel) and of the semi-major axis of 
the orbit of the engulfed planet (lower panel) for Q$^{'}=10^7$.}\label{f:q7}
\end{figure}
\end{center}
\begin{center}
\begin{figure} 
\includegraphics[width=0.4\textwidth]{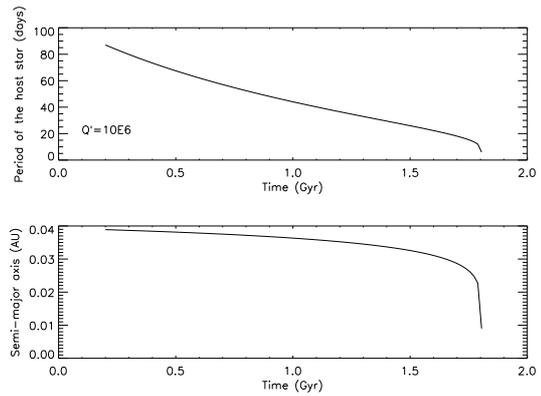}
\caption{As Fig.~\ref{f:q7}, but for Q$^{'}=10^6$.}\label{f:q6}
\end{figure}
\end{center}
We consider that the age of the star is $\approx$ 2 Gyr and that the bulk of the mass 
transfer has not occurred more than 200 Myr ago,  otherwise the acquired angular momentum 
would have been partially lost resulting in a slowdown of the stellar rotation owing to 
its magnetised wind.
For $Q^{'}$=10$^7$ and a stellar rotation period of 6.5 days when the planet reaches its Roche lobe, 
the reverse integration results in an initial orbital distance for the planet of 
$\approx 0.030$~AU and a rotation period of $\approx 20$ days (cf. Fig.~\ref{f:q7}). 
On the other hand, if we adopt $Q^{'}$=10$^6$, the system evolution is so fast and 
tides are so effective that the initial rotation period of the star becomes unrealistically long 
before $\approx 1$~Gyr (cf. Fig.~\ref{f:q6}). This is an indication that, despite all 
the uncertainties involved in our knowledge of the tidal interaction quality factor, 
 $Q^{\prime} \approx 10^7$ seems more realistic. 
In comparison to the estimates by \citet{fuhrmann15}, we find that the approach of 
the planet to the star under the action of tides is accompanied by a significant 
spin up of the star itself as a result of the transfer of angular momentum from  the 
orbit to the stellar spin. Therefore, in our model the initial rotation of the 
star is found  to be rather slow, if tides acted since the system had $\approx 0.2$~Gyr. 
We might speculate that the planet was not initially at $\approx 0.030$~AU; but it 
was brought there through the Kozai mechanism because of the presence of the external 
massive planet/ brown dwarf companion of GJ504. Once this configuration was reached 
tidal forces did their job, causing an orbital decay in approximately 
$1.8-2.0$~Gyr. When the planet reached its Roche lobe, conservative mass 
transfer could have further decreased the 
rotation period of the star from $\approx 7$ days down to a few days. 
We can account for a faster initial stellar rotation at 0.2~Gyr by reducing the 
mass of the planet, its initial distance, and playing with the poorly known tidal 
parameter $Q^{\prime}$. In view of these uncertainties, we regard our computed models 
as simply illustrative and do not attempt a full exploration of the initial parameter 
space to account for the currently fast stellar rotation.

It is interesting to note that, according to our scenario in which the planet is initially based
at a distance of 0.03~AU with a mass of 2.7 M$_{j}$, the location of this hot Jupiter 
falls within the upper envelope of the distributions of planetary masses versus orbital 
period.


\section{Concluding remarks}
In this paper we have presented a comprehensive scrutiny of the fundamental properties for GJ 504. The star hosts 
a substellar companion located at $\approx$ 44 AU, which has been claimed to be one of the lowest mass objects ever observed 
via direct imaging techniques (\citealt{kuzuhara13}). However, this conclusion relied on a young age for GJ 504, namely less than $\sim$ 200 Myr, as indicated by the rotational period of P=3.33 days. Conversely, \cite{fuhrmann15}, thanks to high-resolution spectroscopic observations, found evidence for a much older age ($\approx$ 3 - 6 Gyr), 
suggesting that the substellar companion is actually a brown dwarf rather than a giant planet. In order to reconcile this old age, coming from isochrones and spectroscopy, 
with the indications of youth given by rotational and activity properties, these authors conceived the possibility of a merging event. The star GJ 504 might have engulfed a planetary-mass object that caused the spin up of rotational velocity along with enhancements in the chromospheric activity level.
To shed light on this debated and complex picture, we exploited a strictly differential (line-by-line) analysis of GJ 504 with respect to two reference stars, namely $\iota$ Hor and HIP 84827. 
The role of $\iota$ Hor is particularly critical in this respect because its properties are very well constrained from an independent  tool, i.e. asteroseismic observations
(see \citealt{vauclair08}).
Our results indicate that the surface gravity of GJ 504 is 0.2 $\pm $ 0.07 dex lower than that of the main-sequence star $\iota$ Hor, 
implying a slightly more advanced evolutionary stage for the object. The isochrone comparison provides us with an age range between 1.8 and 3.5 Gyr, which is qualitatively in agreement with Fuhrmann \& Chini results, 
although slightly younger.
Interestingly, the solar Ba abundance also points to a very old age for the system, which is at variance with young stars in clusters and in the field
(\citealt{dorazi09}; \citealt{desidera11}) that are known to be characterised by an extremely high Ba content ([Ba/Fe] up to 0.6 dex). 
To investigate the merging scenario suggested by Fuhrmann \& Chini, we ran a tidal evolution code, assuming an age of roughly $\approx$ 2 Gyr and imposing the condition that the engulfment event could not have occurred more than 200 Myr ago; otherwise the star would have lost the angular momentum and slowed down.
Our tests indicate that a very plausible system architecture would result in an initial configuration of a planetary companion 
(with mass not larger than $\sim$ 3 M$_{J}$) located at 0.03 AU. This is probably because Kozai cycles (due to the presence of the external sub-stellar companion) have caused an inward migration. 
From such a small distance, the low-mass body has been affected by the stellar tides and slowly started to spiral down on the central star. If this were the case, we would expect to reveal planetary remnants such as rocky cores of the now defunct hot Jupiter in the proximity of star GJ 504. 
Current measurements prevent us from investigating this issue, but we plan to have purposely designed observations to detect such a small signal in radial velocity variations.


\begin{acknowledgements}
This work has made extensive use of the SIMBAD, Vizier, and NASA ADS databases.
VD thanks Ivan Ramirez and Tamara Mishenina for very useful discussions and for 
providing information on unpublished material. We thank the anonymous referee for very helpful comments and suggestions.
\end{acknowledgements}

%
%

\end{document}